\documentclass[final,3p,times]{elsarticle}
\usepackage{graphicx}
\usepackage{amssymb}
\usepackage{amsmath}
\usepackage{hyperref}
\hypersetup{
hypertexnames = false,
colorlinks = true,
urlcolor = blue,
linkcolor = blue,
citecolor = blue
}

\journal{Computer Physics Communications}

\begin{document}

\begin{frontmatter}

\title{Spin-1 spin-orbit- and Rabi-coupled Bose-Einstein condensate solver}

\author[bdu]{Rajamanickam Ravisankar}
\ead{ravicpc2012@bdu.ac.in}

\author[scl]{Du\v san Vudragovi\'c}
\ead{dusan@ipb.ac.rs}

\author[bdu,bdu1]{Paulsamy Muruganandam}
\ead{anand@bdu.ac.in}

\author[scl]{Antun Bala\v{z}}
\ead{antun@ipb.ac.rs}

\author[ift]{Sadhan K. Adhikari\corref{author}}
\ead{sk.adhikari@unesp.br}

\cortext[author]{Corresponding author.}

\address[bdu]{Department of Physics, Bharathidasan University, Palkalaiperur Campus, Tiruchirappalli 620024, Tamilnadu, India}

\address[scl]{Institute of Physics Belgrade, University of Belgrade, Pregrevica 118, 11080 Belgrade, Serbia}

\address[bdu1]{Department of Medical Physics, Bharathidasan University, Palkalaiperur Campus, Tiruchirappalli 620024, Tamilnadu, India}
 
\address[ift]{Instituto de F\'{\i}sica Te\'{o}rica, UNESP -- Universidade Estadual Paulista, 01.140-70 S\~{a}o Paulo, S\~{a}o Paulo, Brazil}

\begin{abstract}
We present OpenMP versions of FORTRAN programs for solving the Gross-Pitaevskii equation for a harmonically trapped three-component spin-1 spinor Bose-Einstein condensate (BEC) in one (1D) and two (2D) spatial dimensions with or without spin-orbit (SO) and Rabi couplings. Several different forms of SO coupling are included in the programs. We use the split-step Crank-Nicolson discretization for imaginary- and real-time propagation to calculate stationary states and BEC dynamics, respectively. The imaginary-time propagation programs calculate the lowest-energy stationary state. The real-time propagation programs can be used to study the dynamics. The simulation input parameters are provided at the beginning of each program. The programs propagate the condensate wave function and calculate several relevant physical quantities. Outputs of the programs include the wave function, energy, root-mean-square sizes, different density profiles (linear density for the 1D program, linear and surface densities for the 2D program). The imaginary- or real-time propagation can start with an analytic wave function or a pre-calculated numerical wave function. The imaginary-time propagation usually starts with an analytic wave function, while the real-time propagation is often initiated with the previously calculated converged imaginary-time wave function.
\end{abstract}


\end{frontmatter}

\begin{small}
\noindent
{\bf Program summary}
\noindent\\
{\em Program title:} BEC-GP-SPINOR, consisting of: BEC-GP-SPINOR-OMP package, containing programs spin-SO-imre1d-omp.f90 and spin-SO-imre2d-omp.f90, with util.f90.
\noindent\\
{\em CPC Library link to program files:} \href{https://doi.org/10.17632/j3wr4wn946.1}{https://doi.org/10.17632/j3wr4wn946.1}\\
{\em Licensing provisions:} Apache License 2.0\\
{\em Programming language:} OpenMP FORTRAN. The FORTRAN programs are tested with the GNU, Intel, PGI, and Oracle compiler.
\noindent\\
{\em Nature of problem:}
The present Open Multi-Processing (OpenMP) FORTRAN programs solve the time-dependent nonlinear partial differential Gross-Pitaevskii (GP) equation for a trapped spinor Bose-Einstein condensate, with or without spin-orbit coupling, in one and two spatial dimensions.\\
{\em Solution method:}
We employ the split-step Crank-Nicolson rule to discretize the time-dependent GP equation in space and time. The discretized equation is then solved by imaginary- or real-time propagation, employing adequately small space and time steps, to yield the solution of stationary and non-stationary problems, respectively.
\end{small}

\newpage

\section{Introduction}
\label{sec:intro}

Previously published FORTRAN~\cite{bec2009} and C~\cite{bec2012} programs are now
popular tools for studying the properties of a Bose--Einstein condensate (BEC) by
solving the Gross--Pitaevskii (GP) equation and are enjoying widespread use. These
programs have later been extended to the more complex scenario of dipolar
atoms~\cite{dbec2015} and of rotating BECs~\cite{vor-lat}. The
OpenMP~\cite{bec2016,bec2017x} and CUDA~\cite{dbec2016,dbec2016a,dbec2016b} version
of these programs, designed to make these faster and more efficient in multi-core
computers, are also available. 

There has been great interest in the studies of spinor BECs using the GP equation 
after the experimental observation of the same~\cite{exptspinor,exptspinorb}. 
Later, it has been possible to introduce an artificial synthetic spin--orbit (SO) coupling  by
Raman lasers that coherently couple the spin-component
states in a pseudo spin-1/2~\cite{exptso,exptsob} and spin-1~\cite{exsp1} spinor
BEC.
In this paper, we present new OpenMP FORTRAN programs to solve the GP equation 
for a three-component spin-1 spinor quasi-one-dimensional (quasi-1D) and
quasi-two-dimensional (quasi-2D) BECs~\cite{quasi12d} with~\cite{thso,thsob} or
without~\cite{thspinor,thspinorb} SO and associated Rabi couplings, based on our
earlier programs~\cite{bec2016,bec2017x}. 
The GP equation for an SO-coupled three-component spin-1 trapped BEC is conveniently solved by the imaginary- and real-time propagation methods. 
We provide combined imaginary- and real-time programs in one and two spatial dimensions. The present imaginary-time programs already involve complex variables and are hence combined together with the real-time programs, requiring complex algebra. The choice of the type of propagation (imaginary-
or real-time) is made through an input parameter.
The imaginary-time approach should be used to solve the GP equation for stationary states. A subsequent study of the non-stationary dynamics of the BEC should be done using the real-time propagation using the imaginary-time wave function as the initial state. 
We use the split-step Crank--Nicolson scheme for solving the GP equation, as in
Refs.~\cite{bec2009,bec2012}. 

In Section~\ref{sec:GPE} we present the GP equation for a spin-1 spinor BEC in a trap. The mean-field model
and a general scheme for its numerical solution are considered for both quasi-1D and quasi-2D traps. 
The details about the computer programs, and their input parameters, output
files, etc.~are given in Section~\ref{sec:details}. The numerical method and results
are given in Section~\ref{sec:numerics}, where we illustrate the results for density and energy by employing the imaginary-time propagation for different interaction strengths (nonlinearities).
 The stability of the density profiles is demonstrated in 
real-time propagation using the corresponding converged solution obtained by the imaginary time propagation as the initial state. 
 Finally, a brief summary is given in Section~\ref{sec:con}. Technical and mathematical details of this investigation are presented in two Appendices. 
 A novel numerical procedure applied in this study is given in \ref{appA}.
 Useful analytic variational and Thomas--Fermi (TF) approximations are developed in the Supplementary material.

\section{The Gross-Pitaevskii equation for a spin-1 condensate}
\label{sec:GPE}

In the mean-field approximation a quasi-1D or quasi-2D SO and Rabi coupled  spin-1 $(F=1)$ BEC is described by the 
 following set of three coupled GP equations, for $N$ atoms of mass $\widetilde m$ each,
in dimensionless form, for the spin components
$F_z = \pm 1, 0$~\cite{thspinor,thspinorb,GA}
\begin{align}
\label{EQ1} 
i {\partial_t \psi_{\pm 1}({\mathbf{r}})}&= \left[-\frac{1}{2}\nabla^2+V({\mathbf{r}})+c_0 \rho+{c_2}
\left(\rho_{\pm 1} -\rho_{\mp 1} +\rho_0\right) \right] \psi_{\pm 1}({\mathbf{r}})
+ \left\{c_2 \psi_0^2({\mathbf{r}})\psi_{\mp 1}^{*}({\mathbf{r}})\right\} + \frac{\Omega}{\sqrt 2}\psi_{0}({\mathbf{r}})+\gamma f_{\pm 1} \, , 
\\ \label{EQ2}
i {\partial_t \psi_{0}({\mathbf{r}})}&=\left[-\frac{1}{2}\nabla^2+V({\mathbf{r}})+c_0 \rho+{c_2}
\left(\rho_{+ 1}+\rho_{- 1}\right)\right] \psi_{0}({\mathbf{r}})
+ \left \{ 2c_2 \psi_{+1}({\mathbf{r}})\psi_{-1}({\mathbf{r}})\psi_{0}^{*} ({\mathbf{r}})\right\}+ \frac{\Omega}{\sqrt 2} \sum_{j=+1,-1} \psi_{j}({\mathbf{r}})
+\gamma g  \, , 
\end{align}
where $\rho_j = | \psi_j |^2$ are the densities of components $j= \pm 1 , 0$, and $\rho ({\mathbf{r}})= \sum \rho_j({\mathbf{r}})$ is the total density, $V({\mathbf{r}})$ is the confining trap,  $\partial_t $ ($\partial_{\mathbf{r}}\equiv \{\partial_x,\partial_y, \partial_z  \}$) is the partial time (space) derivative,  and  $\Omega$ ($\gamma$) is the strength of Rabi (SO) coupling. The SO coupling is  a space derivative coupling described by the functions $f$ and $g$, the details of which are given 
below. For brevity, the time dependence of the wave functions is not explicitly shown  in Eqs.~(\ref{EQ1}) and (\ref{EQ2}).
In 1D, ${\mathbf{r}}= x, \nabla^2 = \partial_x^2 = \partial^2/\partial x^2$, 
in 2D, $ {\mathbf{r}}= \{ x,y \}, \nabla^2 = \partial_x^2 +\partial_y^2$, and 
in 3D, $ {\mathbf{r}}= \{ x,y, z \}, \nabla^2 = \partial_x^2 +\partial_y^2+\partial _z^2$. In 3D, distances are expressed in units of the harmonic oscillator length $l\equiv \sqrt{\hbar/\widetilde m\omega}$, density 
$\rho_j$ in units of $l^{-3}$ and time in units of $\omega^{-1}$, where
$\omega=\omega_x$ is the $x$-axis trapping frequency. The potential is $V({\mathbf{r}})= (x^2+\kappa^2 y^2+ \beta^2 z^2 )/2$,
where the trap aspect ratios are $\kappa=\omega_y/\omega$ and $\beta=\omega_z/\omega$. The dimensionless
nonlinearities are $c_i= 4\pi N {\mathcal{A}}_i$, $i=0,2$, where ${\mathcal{A}}_0 = (a_0+2a_2)/3l,{\mathcal{A}}_2 = (a_2-a_0)/3l$, with $a_0$ and
$a_2$ being the scattering lengths in the total spin channels 0 and 2,
respectively. For a pancake-shaped trap, with the strong trapping in $z$
direction $(\beta \gg 1, \kappa)$, a set of quasi-2D~\cite{quasi12d} equations can be obtained
with $ c_i= 2\sqrt{2\pi\beta } N {\mathcal{A}}_i,$\,   with $ V({\mathbf{r}})= (x^2+\kappa^2 y^2)/2$. For a cigar-shaped trap, with the strong
trapping in $y$ and $z$ directions $(\beta, \kappa \gg 1)$, a set of
quasi-1D~\cite{quasi12d} equations can be obtained with $ c_i= 2 \sqrt{\kappa\beta}N {\mathcal{A}}_i,$\, where 
$ V({\mathbf{r}})= x^2/2$.
In the following we will take $\kappa=1$ and $V({\mathbf{r}})= {\mathbf{r}}^2/2$ in both 1D and 2D. In the programs the parameter $\kappa$ in the potential is set to unity, but a different value can be introduced easily if needed.

In the presence of the SO coupling \cite{thso,thsob}, we consider below the SO-coupling contributions $\gamma f_{\pm 1}$ and 
 $\gamma g $ of Eqs.~(\ref{EQ1}) and (\ref{EQ2}) in different cases. In 1D we consider three possible SO couplings in the Hamiltonian: $\gamma p_x \Sigma_x$,
$\gamma p_x \Sigma_y$, and $\gamma p_x \Sigma_z$, where $p_x = -i \partial_x$ is the momentum operator and $\Sigma_x$,
$\Sigma_y$ and $\Sigma_z$ are the irreducible representations of the $x, y$ and $z$ components of the spin-1
matrix $\Sigma$, with components
\begin{align}\label{smat}
\Sigma_x= \frac{1}{\sqrt 2} \left( \begin{array}
 {ccccc}
0 & 1 & 0\\
1 & 0 & 1\\
0 & 1 & 0 \end{array} \right)\, , \quad 
\Sigma_y= \frac{i}{\sqrt 2} \left( \begin{array}
 {ccccc}
0 & -1 & 0\\
1 & 0 & -1\\
0 & 1 & 0 \end{array} \right)\, , \quad 
\Sigma_z= \left( \begin{array}
 {ccccc}
1 & 0 & 0\\
0 & 0 & 0\\
0 & 0 & -1 \end{array} \right)\, .
\end{align}
For the SO coupling $\gamma p_x \Sigma_x$ \cite{lplga,ska} in 1D, the SO coupling terms in Eqs. ~(\ref{EQ1}), (\ref{EQ2}) are $\gamma f_{\pm 1}= - i\widetilde \gamma \partial_x \psi_0 ({\bf r})$ and 
$\gamma g = - {i\widetilde \gamma} \Big[\partial_x \psi_{+1}({\bf r}) 
+\partial_x \psi_{-1}({\bf r}) \Big],$ respectively,
where $\widetilde \gamma = \gamma/\sqrt{2}$.
 For the SO coupling $\gamma p_x \Sigma_y$ they are $ \gamma f_{\pm 1}= \mp i\widetilde \gamma \partial_x \psi_0 ({\bf r})$ and 
$ \gamma g= {i\widetilde \gamma} \Big[\partial_x \psi_{+1}({\bf r}) 
-\partial_x \psi_{-1}({\bf r}) \Big],$ respectively.
 For the SO coupling $\gamma p_x \Sigma_z$ they are \cite{ska} 
$\gamma f_{\pm 1}= \mp {i\gamma} \partial_x \psi_{\pm 1} ({\bf r}) $ and 
$\gamma g= 0$,
respectively.

In 2D we consider the general SO coupling term in the form $\gamma(\eta p_y \Sigma_x - p_x \Sigma_y)$, where $\eta=1,-1$ and 0 for Rashba \cite{SOras}, Dresselhaus \cite{SOdre} and an equal mixture of Rashba and Dresselhaus SO couplings. 
 In Eqs. ~(\ref{EQ1}), (\ref{EQ2}), the Rashba, Dresselhaus and an equal mixture of Rashba and Dresselhaus coupling terms in 2D are $\gamma f_{\pm 1}= - {i\widetilde \gamma}\Big[\eta \partial_y \psi_0 ({\bf r}) \pm i \partial_x \psi_0 ({\bf r}) \Big]$ and 
$\gamma g= - {i\widetilde \gamma} \Big[-i \partial_x \psi_{+1}({\bf r}) 
+i\partial_x \psi_{-1}({\bf r}) + \eta\partial_y \psi_{+1}({\bf r}) +
\eta \partial_y \psi_{-1}({\bf r}) 
\Big] $. 
 
The normalization and magnetization ($m$) conditions are given by
\begin{align}\label{noma}
 \int \rho({\bf r})\, d{\bf r}=1\, , \quad \text{and} \quad \int \Big[\rho_{+1}({\bf r})
-\rho_{-1}({\bf r})\Big] \, d{\bf r}=m\, .
\end{align}
Condition (\ref{noma}) is useful to solve the problem for a fixed normalization when magnetization $m$ along $z$ direction is conserved, e.g., when the Hamiltonian commutes with spin-matrix $\Sigma_z$. However, in the presence of an SO coupling, that does not commute with $\Sigma_z$, magnetization is not conserved due to spin mixing dynamics involving spin-up and down states. In that case, time propagation is performed by imposing only the condition 
of conservation of normalization without fixing the magnetization during time propagation \cite{bao} and it 
leads to the result for the dynamically stable
stationary state with a magnetization determined by the parameters of the model, which could often be zero. In this
context, it should be noted that in experiments it is not possible to fix a preassigned value to magnetization, which is
not a constant of motion.

The energy functional of the system is \cite{thspinor,thspinorb}
\begin{align}\label{energy}
E=&\frac{1}{2} \int d {\bf r}\, \Bigg\{\sum_j |\nabla _{\bf r}\psi_j|^2+2V({\bf r})\rho + c_0 \rho^2
\nonumber \\ &
+ c_2\left[ \rho_{+1}^2+ \rho_{-1}^2 +2 \left( \rho_{+1}\rho_0+ \rho_{-1}\rho_0- \rho_{+1}\rho_{-1}+\psi_{-1}^*\psi_0^2\psi_{+1}^*+
\psi_{-1}{\psi_0^*}^2\psi_{+1}
\right) 
\right] \nonumber \\ &
+2 \frac{\Omega}{\sqrt 2}\Big[ (\psi^*_{+1}+ \psi^*_{-1})\psi _0+ \psi_0^* (\psi_{+1}+ \psi_{-1}) \Big] 
 + 2
\gamma \Big[ 
\psi_{+1}^*f_{+1}+ \psi_{-1}^*f_{-1} + \psi_0^* g 
 \Big] \Bigg\}.
\end{align}

\section{Details about the programs}
\label{sec:details}

We use the split time step Crank-Nicolson discretization rule for solving the GP equations (\ref{EQ1}) and (\ref{EQ2}), including the appropriate SO and Rabi couplings with strengths $\gamma$ and $\Omega$, respectively. This approach has been elaborated in details in Ref.~\cite{bec2009}. An initial (known) wave function at time $t$ is used to calculate the wave function at time $t+\Delta$, after a small time step $\Delta$. The advantage of this approach lies in the fact that different terms on the right-hand-side of Eqs. (\ref{EQ1}) and (\ref{EQ2})
can be treated successively. For example, the spatial derivative term involving $\nabla _{\bf r}$ can be treated independently of the nonlinear interaction terms and also of the SO and Rabi coupling terms. The terms in the square brackets of Eqs. (\ref{EQ1}) and (\ref{EQ2}) can be treated in a routined way elaborated in 
 Ref.~\cite{bec2009}. The terms in the curly brackets and those proportional to $\Omega$ in Eqs. (\ref{EQ1}) and (\ref{EQ2}) need special 
attention and are treated as in \ref{appA}. Finally, the $\gamma$-dependent SO coupling terms only involve first order space derivatives and are treated in a routined fashion. 

\begin{table}[!t]
\caption{Different modules of the 1D and 2D programs and their usage }
\label{tab1}
\begin{tabular}{ll|ll }
\hline
\hline
 & Module name& Type & Usage \\ 
\hline
 & IMRE1D & MAIN & Main program \\ 
 & INITIALIZE & Subroutine & Calculates or reads the initial function \\
 & CALC$\_$TRAP & Subroutine & Calculates the confining trap \\
 & COEF & Subroutine & Calculates the coefficients of the Crank-Nicolson method \\
 & SO & Subroutine & Propagates the spin-orbit coupling term in time \\ 
1D & LU & Subroutine & Crank-Nicolson time propagation \\
 & HERM & Subroutine & Propagates the off-diagonal terms of the GP equation \\
 & & & and the Rabi coupling term in time\\
 & CALCNU & Subroutine & Propagates the diagonal parts of the GP equation in time \\
 & LENGTH & Subroutine & Calculates the length of the condensate \\
 & RENORM & Subroutine & Fixes the normalization and magnetization \\
 & ENERGY & Subroutine & Calculates the energy of the condensate \\
 & SIMP & Function & Performs integration by Simpson's rule \\
\hline
 & IMRE2D & MAIN & Main program \\ 
 & INITIALIZE & Subroutine & Calculates or reads the initial function \\
 & CALC$\_$TRAP & Subroutine & Calculates the confining trap \\
 & COEF & Subroutine & Calculates the coefficients of the Crank-Nicolson method \\
 & SO & Subroutine & Propagates the spin-orbit coupling term in time \\ 
 & LUX & Subroutine & Crank-Nicolson time propagation in $x$ variable \\
2D & LUY & Subroutine & Crank-Nicolson time propagation in $y$ variable \\
 & HERM & Subroutine & Propagates the off-diagonal terms of the GP equation \\
 & & & and the Rabi coupling term in time\\
 & CALCNU & Subroutine & Propagates the diagonal parts of the GP equation in time \\
 & INTEGRATE & Function & Performs double integration in $x$ and $y$ \\
 & RADIUS & Subroutine & Calculates the radius of the condensate \\
 & RENORM & Subroutine & Fixes the normalization and magnetization \\
 & ENERGY & Subroutine & Calculates the energy of the condensate \\
 & SIMP & Function & Performs integration by Simpson's rule \\
\hline
\end{tabular}
\end{table}

\begin{table}[!t]
\caption{Name and description of input parameters and output files }
\label{tab2}
\begin{tabular}{lll }
\hline
 & Name& Description  \\  
\hline
 & NSTP, NPAS, NRUN& Number of time iterations, \\
 &    & NSTP = 0 reads initial function,\\
 &     &  NSTP $>0$ calculates initial function\\
 & N, NX, NY &  Number of space integration points \\
 & NTHREADS  & Number of threads used  \\
 &           & NTHREADS = 0 uses all threads  \\   
Input & C$\_$0, C$\_$2 & Nonlinear input parameters ($c_0,c_2$)\\ 
 & OPT$\_$SO &  Selects the type of SO coupling     \\
 & DX, DY& Space discretization steps  \\
 &   OPT$\_$PROP  & Selects the type of time propagation \\
  &               &  imaginary (=1) or real (=2) time \\
  &     OPT$\_$ST & Selects initial function in 2D     \\
 & MAG$\_$0 & Magnetization  ($m$)\\
 & GAM0  & Strength of SO coupling ($\gamma$)  \\
 & OMEGA0 & Strength of Rabi coupling   ($\Omega$)\\
\hline
 &\texttt{im-out.txt} & Input parameters, energy and size  \\ 
 &  \texttt{im-den-<desc>.txt} & \texttt{<desc>=ini} initial $\&$ \texttt{=fin} final density  \\
Output  &  \texttt{im-phase.txt} & phase of a 2D wave function \\
  &  \texttt{im-wave-fun-<desc>.txt} & \texttt{<desc>=ini} initial $\&$ \texttt{=fin} final wave function   \\ 
 &  \texttt{im-den-rad-<desc>.txt} & \texttt{<desc>=ini} initial $\&$ \texttt{=fin} final radial density (2D)  \\
\hline
\end{tabular}
\end{table}

The presented programs are straightforward modifications of the basic programs published in Refs.~\cite{bec2009,bec2012}. The three components of the wave function are accommodated by 
introducing a new index ``L'' in addition to the space indices in the corresponding arrays, i.e., the wave-function components are represented by CP(L,I,J) in 2D and CP(L,I) in 1D, where L=1,2,3 stands for the spin components $j=+1$, 0, and $-1$, respectively, and I and J denote discretized space points. The time propagation with respect to different terms in Eqs.~(\ref{EQ1}) and (\ref{EQ2}) are dealt with in different subroutines. The kinetic energy term ($\nabla^2/2$) is 
treated using the Crank-Nicolson discretization in subroutines COEF, LUX, and LUY in 2D, and in 1D in subroutines COEF and LU, as in Refs.~\cite{bec2009,bec2012}. The potential term and the diagonal part of the nonlinear terms in square brackets, proportional to $c_0$ and $c_2$, are treated in the subroutine 
CALCNU. The off-diagonal part of the nonlinear terms in curly brackets in Eqs.~(\ref{EQ1}) and (\ref{EQ2}), 
explicitly considered in Eq.~(\ref{eq:nu}), is treated in the subroutine HERM, while the different SO coupling terms 
are treated in the subroutine SO. The conservation of the normalization and magnetization, 
as defined by Eqs.~(\ref{DJ}) and (\ref{DJ2}), is implemented in the subroutine RENORM. The energies are calculated in the subroutine ENERGY and expectation values of the condensate's cloud sizes are calculated in the subroutine 
RADIUS in 2D and LENGTH in 1D. The different modules (Subroutines and Functions) of the programs in 1D
and 2D and their respective usage are presented in Table \ref{tab1}. 

A description of the input parameters together with the output files with description are given in Table \ref{tab2}. Most of the parameters have the same meaning as in our previously published programs and the reader can refer to Ref.~\cite{bec2017x} for details.
For an efficient performance on computers with multiple CPU cores, the programs have been parallelized using the OpenMP library. The number of threads (CPU cores) to be used is declared by the parameter NTHREADS, which should be equal to or less than the total number of available threads. If NTHREADS is set to zero, then all available CPU cores will be used. 
 The parameters NSTP, NPAS, and NRUN denote different numbers of time iterations, the total number of iterations being the sum of these. If NSTP is different from zero, then the program starts the time propagation using an analytic initial function defined in the subroutine INITIALIZE. If NSTP is zero, the program reads an initial wave function for the calculation from input files, i.e., from the previously calculated files named 
\texttt{<code>-wave-fun-fin.txt}. 
In this way one can perform the imaginary- or real-time propagation with a pre-calculated wave function.
 The supplied programs use the pre-defined value \texttt{NSTP = 10} and use an analytic wave function as the initial state. 
 When using a pre-calculated wave function by setting NSTP to zero, the number of space grid points N (1D) 
and NX, NY (2D) employed previously should match exactly the number of points used in the current program. The supplied programs assume equal numbers of space step points 
in both imaginary- and real-time propagation. The parameter OPT$\_$SO selects the type of SO coupling. In 1D 
OPT$\_$SO =1,2,3 uses the SO couplings $\gamma p_x \Sigma_x$, $\gamma p_x \Sigma_y$, and $\gamma p_x \Sigma_z$, respectively.
In 2D OPT$\_$SO =1,2, and 3 uses Rashba, Dresselhaus, and an equal mixture of Rashba and Dresselhaus SO coupling, respectively. 
The choice OPT$\_$SO =0 corresponds to no SO coupling. The parameters MAG$\_$0, GAM0 (GAMMA0 in 1D) and OMEGA0 denote magnetization, 
 the strength of SO coupling $\gamma$ and that of Rabi coupling $\Omega$, respectively. The parameter OPT$\_$PROP selects the type of 
time propagation: imaginary (1) and real (2) time.
All input data are conveniently placed at the beginning of each program, as before~\cite{bec2016}. 
After changing the input data in a program a recompilation is required. 
The output files are conveniently named such that their contents can be easily identified, following the naming convention introduced in Ref.~\cite{bec2016}. For example, a file named \texttt{<code>-out.txt}, where \texttt{<code>} denotes imaginary- (im) or real-time (re) propagation, represents the general output file containing input data, space and time steps DX, DY and DT,
nonlinearity $c_0$ and $c_2$, energy, size, etc. The files \texttt{<code>-den-ini.txt} and \texttt{<code>-den-fin.txt} contain the initial and final component and total densities in different columns. In 1D (2D) these densities are functions of one (two) space point(s) placed in the first (and second) column(s) of the respective files. The densities $\rho_{+1}$, $\rho_{0}$, $\rho_{-1}$, and $\rho$ can be found in the successive columns.
 The file \texttt{<code>-den-rad-fin.txt} 
stores
the final linear radial densities $\rho_{+1}$, $\rho_{0}$, $\rho_{-1}$, and $\rho$ for the 2D GP equation in different columns, while the space points are saved in the first column. 
The file \texttt{<code>-wave-fun-fin/ini.txt} contains the final/initial complex wave functions.
For a 2D BEC, the file \texttt{<code>-phase.txt} contains the phases of the component wave functions 
in different columns, since the phase is important for the study of angular momentum of the respective states. 
The printing of some of these files, such as the initial density and wave function, are commented out in the programs, so that the supplied programs do 
not print these.

We provide below the beginning of the 1D program where the parameters are defined so that 
the reader can easily identify the different statements there. The 2D program is quite similar. 
\begin{verbatim}
! Begin selection of input parameters
MODULE COMM_DATA
! SELECT # OF SPACE POINTS N AND # OF TIME ITERATIONS NSTP, NPAS & NRUN
! USE NSTP = 0 TO READ WAVE FUNCTION FILE FROM STDIN: < im-wave-fun-fin.txt
  INTEGER, PARAMETER :: N = 640, NX = N-1, NX2 = N/2
  INTEGER, PARAMETER :: NSTP = 10, NPAS = 1000000, NRUN = 100000
! INTEGER, PARAMETER :: NSTP = 0, NPAS = 1000000, NRUN = 100000 
! Number of OpenMP threads, less than or equal to the maximum available cores.
  INTEGER, PARAMETER  :: NTHREADS = 0  ! NTHREADS = 0 uses all available cores  
  REAL (8), PARAMETER :: Pi = 3.14159265358979D0
 END MODULE COMM_DATA
!******************************************************************
MODULE SPIN_PARS
  USE COMM_DATA, ONLY : PI
!******* SELECT  POLAR (C_2 > 0) OR FERROMAGNETIC (C_2 < 0) BEC ******
! REAL (8), PARAMETER :: C_0 = 241.d0, C_2 =  7.5d0  ! Anti-ferromagnetic
  REAL (8), PARAMETER :: C_0 = 885.d0, C_2 = -4.1d0  ! Ferromagnetic
END MODULE SPIN_PARS
!******************************************************************
MODULE GPE_DATA
  USE COMM_DATA, ONLY : N, Pi
  USE SPIN_PARS, ONLY : C_0, C_2
!***************************************************************************************
!*** SELECT OPTION FOR SO COUPLING AND STRENGTH GAMMA
! INTEGER,PARAMETER :: OPT_SO = 0; REAL (8), PARAMETER :: GAMMA0 = .00D0  ! No SO coupl.  
  INTEGER,PARAMETER :: OPT_SO = 1; REAL (8), PARAMETER :: GAMMA0 = .500D0 ! Sigma_x p_x
! INTEGER,PARAMETER :: OPT_SO = 2; REAL (8), PARAMETER :: GAMMA0 = .500D0 ! Sigma_y p_x
! INTEGER,PARAMETER :: OPT_SO = 3; REAL (8), PARAMETER :: GAMMA0 = .500D0 ! Sigma_z p_x
!***************************************************************************************
  REAL (8), PARAMETER :: DX = 0.05D0  !  SELECT SPACE STEP DX
!!SELECT OPTION  FOR  PROPAGATION: IMAGINARY-TIME or REAL-TIME
  INTEGER, PARAMETER :: OPT_PROP = 1; REAL (8), PARAMETER :: DT = DX*DX*0.10D0 ! IMAG
! INTEGER, PARAMETER :: OPT_PROP = 2; REAL (8), PARAMETER :: DT = DX*DX*0.05D0 ! REAL
!!SELECT PARAMETERS OF MODEL
  REAL (8), PARAMETER :: MAG_0 = .400000D0,  ACCUR=1.D-6  ! Magnetization
  REAL (8), PARAMETER :: OMEGA0 = .000D0                  ! Rabi and SO coupling
!***************************************************************************************
! End selection of input parameters
\end{verbatim}

 Below we provide a sample output file re-out.txt for the 2D program for the readers to 
familiarize. 
\begin{verbatim}
     REAL-TIME PROPAGATION
     # of threads   =          16
 
     RASHBA SO coupling, GAMMA =   0.500000000000000
     RABI coupling Omega =   0.000000000000000E+000
     Nonlinearity C_0 = 482.000000, C_2 = 15.000000
     OPT_ST =   0.750000000000000
 
     Space and time steps: DX =  0.10000,  DY =  0.10000, DT =  0.2500E-03
     # of space steps: NX =   161,  NY = 161
     # of time steps:  NSTP = 0, NPAS = 80000, NRUN = 10000
 
            ------------------------------------------------------------
                  RAD(1)      RAD(2)      RAD(3)     Energy      MAG
            ------------------------------------------------------------
NSTP iter.:       3.469       2.194       3.469      8.1969      0.0000
NPAS iter.:       3.468       2.193       3.468      8.1969     -0.0000
NRUN iter.:       3.469       2.195       3.469      8.1969     -0.0000
            ------------------------------------------------------------
 
 Clock Time:     129 seconds
   CPU Time:    2048 seconds
\end{verbatim}

\begin{figure}[!b]
\centering
\includegraphics[width=.6\linewidth]{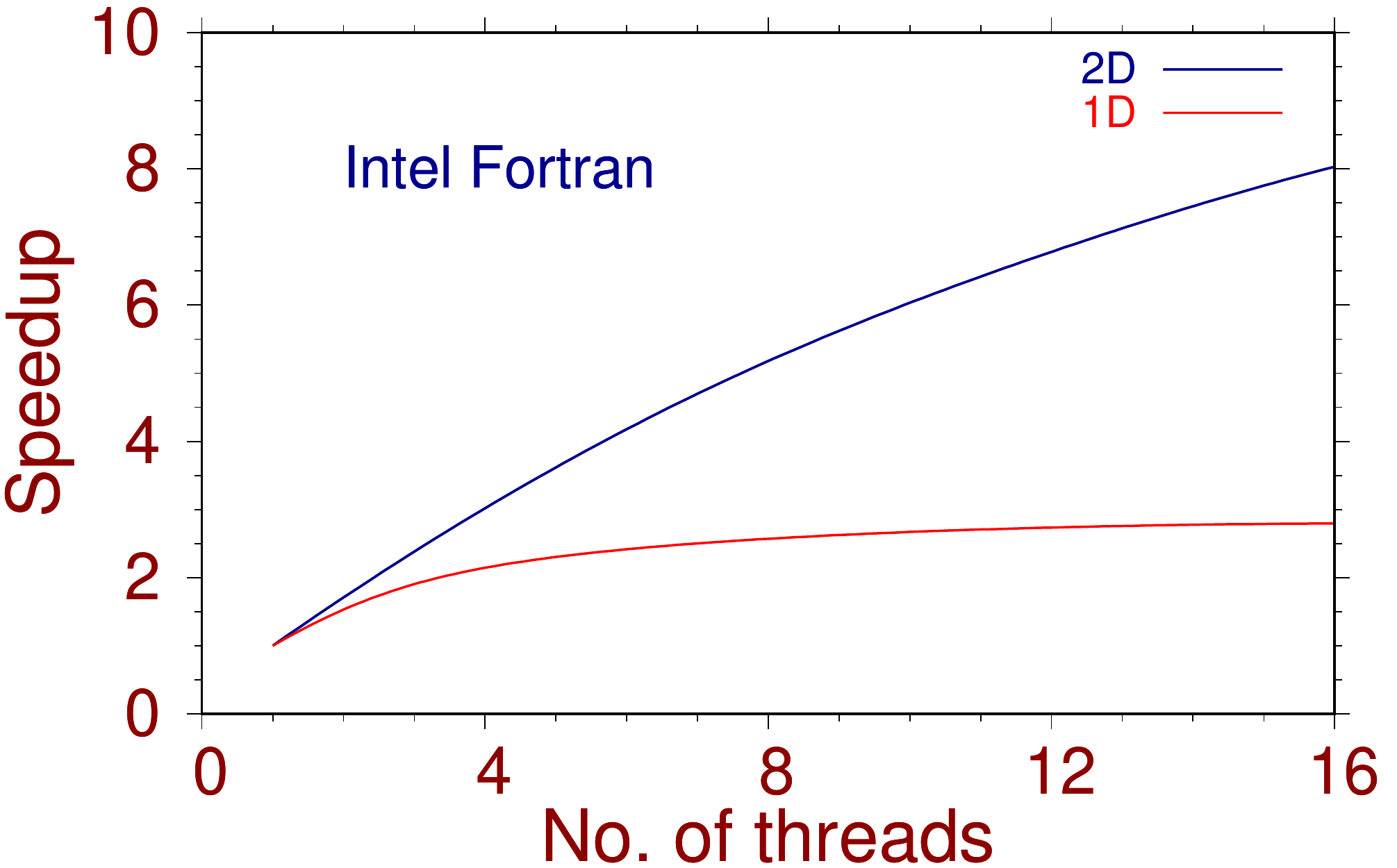}
\caption{Speedup of execution on a 20-core machine with two Intel Xeon E5-2650 processors ($2\times 10$ CPU cores)
versus the number of threads used, for a quasi-1D and quasi-2D BEC for a 
typical run. The size of the space grid in 1D is 1000 and in 2D it is 200$\times$200. The programs were compiled using the Intel Fortran compiler v.~19.0.4.243.}
\label{fig1}
\end{figure}

Another crucial aspect for the execution of imaginary-time propagation to find the lowest-energy ground state is a proper choice of initial state with right symmetry property as the final state. Different type of states can be obtained for different sets of parameters. 
Without SO coupling, the solution is of the Gaussian type and a Gaussian function should be chosen as the initial state.
For small Rashba or Dresselhaus SO-coupling strength $\gamma$ $(\gamma \lessapprox 0.75)$, the lowest-energy circularly-symmetric state of the three components of a quasi-2D SO-coupled anti-ferromagnetic (polar) spin-1 BEC is of the $(-1,0,+1)$ or $(+1,0,-1)$ type, where the numbers in the parenthesis represent the angular momentum of the vortices in the centre of the components $j=+1,0,-1$, respectively, with the negative sign representing an anti-vortex. 
For a ferromagnetic spin-1 quasi-2D BEC these states are of the type $(0,\pm 1,\pm 2)$ for Rashba and Dresselhaus SO-couplings, respectively, in agreement with the consideration of Ref. \cite{kita}. For an equal mixture of Rashba and Dresselhaus SO couplings, for small $\gamma$, the lowest-energy 
ferromagnetic BEC states are of the Gaussian type without any vortices; the anti-ferromagnetic BEC states are of the stripe type with periodic 1D modulation in density.
For an efficient computation, the vortices or anti-vortices are introduced in the initial wave functions when required.
The vortex of angular momentum $L$ is imprinted by taking the initial state as a Gaussian 
multiplied by the factor $(x+iy)^L$, and an anti-vortex by the factor $(x-iy)^L$. 
 For large SO-coupling strength $\gamma$ $(\gamma \gtrapprox 0.75)$, the density of the lowest-energy ground state of the three components of an SO-coupled spinor BEC exhibits different patterns (not considered in this paper). The initial states in the 2D program has to be chosen accordingly. A stripe pattern is generated by multiplying an initial Gaussian state by $\sin(\gamma x)$ and $\cos(\gamma x)$. The user can change the initial state for a SO-coupled BEC by choosing the value of the parameter OPT$\_$ST. For $\gamma <$ OPT$\_$ST the states of type $(\mp 1,0,\pm 1)$ and $(0,\pm 1,\pm 2)$ are chosen in anti-ferromagnetic and ferromagnetic phases, the upper (lower) sign corresponds to Rashba (Dresselhaus) SO coupling. 
In the anti-ferromagnetic phase, for $\gamma >$ OPT$\_$ST, the stripe states are chosen as the initial state. 
However, 
to reproduce the results reported in this paper there is no need to change the parameter OPT$\_$ST.
For an equal mixture of Rashba and Dresselhaus SO couplings Gaussian-type initial states are appropriate for the ferromagnetic phase and stripe states for the 
anti-ferromagnetic phase. 
 
If the imaginary-time propagation is performed, the programs run either by using an initial 
analytic input function (if \texttt{NSTP} is not set to zero), or by employing a pre-calculated wave function (if \texttt{NSTP} is set to zero). The real-time propagation can successfully work only if initialized with a meaningful wave function, usually assuming that \texttt{NSTP = 0} is set, and that the program will read a pre-calculated wave function by the earlier performed imaginary-time propagation. 
The calculation is essentially done within the \texttt{NPAS} time iteration loop. Another NSTP time iteration 
is accommodated to verify if the results converged by comparing the energies and sizes after NPAS and NSTP iterations. 
The source programs spin-SO-imre1d-omp.f90 (1D) and spin-SO-imre2d-omp.f90 (2D) are located in the directory \texttt{src} within the corresponding package directory \texttt{BEC-GP-SPINOR-OMP}.
 A \texttt{README.md} file, included in the corresponding root directory, explains the procedure to compile and run the programs in more detail using a makefile. 
In the beginning of each program the compilation commands are given for GNU, Intel, PGI, and Oracle (former Sun) Fortran compilers. They can be compiled by the \texttt{make} command using the provided \texttt{makefile} in the corresponding package root directory. Otherwise, they can be compiled by the commands given at the beginning of the programs using Intel, GNU, PGI, and 
Oracle FORTRAN compilers.
The examples of produced output files can be found in the directory \texttt{output}, although some large density files are omitted, to reduce the software package size. 

We conclude this section demonstrating the efficiency of our OpenMP parallelization scheme using the Intel compiler on a machine
with $2\times 10$ CPU coresin Fig.~\ref{fig1}, where we plot the speedup versus the number of threads. The speedup for $n$ threads is defined as the ratio of clock time 
for a single thread to that for $n$ threads. From Fig. \ref{fig1} we find that the clock time reduces with the increase of number of threads, thus making the execution more efficient on a multi-core machine. 

\section{Numerical results}
\label{sec:numerics}

 All calculations reported below were performed with the predefined space and time steps DX and DT in the 
programs: in 1D DX = 0.05, DT = DX*DX*0.1 (imaginary time) and DT = DX*DX*0.05 (real time); in 2D
 DX = 0.1, DT = DX*DX*0.1 (imaginary time) and DT = DX*DX*0.025 (real time). To increase the accuracy of calculation, 
the space step(s) DX and DY should be reduced and the total number of space discretization points N, NX, and NY increased proportionally.

The parameters of the GP equation $c_0$ and $c_2$ are taken from the following realistic experimental situations. 
 For the ferromagnetic BEC the quasi-1D trap parameters are $l=2.41927$ $\mu$m, $l_{yz}=0.54$ $\mu$m
and we use the following parameters of $^{87}$Rb atoms: $N=10,000, a_0=101.8a_B, a_2=100.4a_B,$ where $a_B$ 
is the Bohr radius. Consequently, $c_0\equiv 2N(a_0+2a_2)l/3l_{yz}^2 \approx 885$ and $c_2\equiv 2N(a_2-a_0)l/3l_{yz}^2\approx -4.1$. 
For the quasi-1D anti-ferromagnetic 
BEC we use the trap parameters $l=4.7$ $\mu$m, $l_{yz}=1.05$ $\mu$m
following parameters of $^{23}$Na atoms: $N=10,000, a_0=50.00a_B, a_2=55.01a_B$. Consequently, $c_0\approx 241$ and $c_2\approx 7.5$. 
For the quasi-2D ferromagnetic 
BEC we use the following parameters of $^{87}$Rb atoms: $N=100,000, a_0=101.8a_B, a_2=100.4a_B,$ \cite{kokk} $l_{z}=2.0157$ $\mu$m. Consequently, 
$c_0\equiv 2N\sqrt{2\pi}(a_0+2a_2)/3l_{z} \approx 1327.5$ and $c_2\equiv 2N\sqrt{2\pi}(a_2-a_0)/3l_{z} \approx -6.15$. 
For the quasi-2D anti-ferromagnetic 
BEC we use the following parameters of $^{23}$Na atoms: $N=100,000, a_0=50.00a_B, a_2=55.01a_B$, \cite{naa} $l_{z}=2.9369$ $\mu$m. Consequently, $c_0 \approx 482$ and $c_2 \approx 15$. 
 
Although we will calculate the lowest-energy ground state by imaginary-time propagation,
it is possible that in some cases {for larger values of SO coupling strength $\gamma$ (not considered in this paper) the imaginary-time approach may converge to a nearby excited state instead of the lowest-energy ground state for certain initial states. The symmetry, such as parity, of the initial state plays a vital role. An even-parity (odd-parity) initial state will find the lowest-energy state 
with even (odd) parity. For small $\gamma$ there are only a few possibilities of symmetry and this problem does not appear for the results reported in this paper. But for larger $\gamma$, 
and especially in the quasi-2D case, there are states 
with many possibilities of symmetry and it may not be easy to know, a priori, the symmetry of the lowest-energy state.} Hence, for large $\gamma$, it is advised to repeat 
the calculation with different initial states, so as to be sure that the converged state is indeed the lowest-energy ground state. In fact, any numerically computed final wave function, obtained with the same number of space points, can be used as the initial state for a new calculation. 

\begin{table}[!t]
\caption{Convergence of energies of harmonically trapped 1D and 2D spin-1 ferromagnetic (ferro) and anti-ferromagnetic (polar) BECs with change of space steps DX and DY. The parameters in 1D: $c_0= 885$, $c_2 = -4.1$ (ferro) and $c_0= 241$, $c_2 = 7.5$ (polar). The parameters in 2D: $c_0= 1327.5$, $c_2 = -6.15$ (ferro), $c_0= 482$, $c_2 = 15$ (polar). The SO-coupled results in 1D refer to the coupling $\gamma p_x \Sigma_x$ with $\gamma=0.5$ and those in 2D correspond to the Rashba coupling with $\gamma =0.5$. The parameters considered here are appropriate for a Rb and Na BEC with trap parameters of Ref.~\cite{bao}. }
\label{tab3}
\centering
\small
\begin{tabular}{ccccccccc}
\hline
\hline
 & 1D (ferro)& 1D (ferro) & 1D (polar)& 1D (polar)& 2D (ferro)& 2D (ferro) & 2D (polar) &2D (polar)\\ 
DX (=DY) & energy& $\%$ error &energy&$\%$ error &energy&$\%$ error & energy & $\%$ error \\ 
\hline
0.4 &{ 36.01167}& 0.00058& { 15.12364}&{0.00066} & {13.59737}&{0.00176} & { 8.19636}&{0.00660} \\ 
0.2 & { 36.01146}& 0& { 15.12354}& {0} &{13.59759}& 0.00015 & { 8.19687}& {0.00037}\\
0.1 & { 36.01146}& 0& { 15.12354}&0 & {13.59761}& 0 & { 8.19690}& 0\\
$< 0.05$ &{ 36.01146}& 0&{ 15.12354}& 0& {13.59761} & 0 &{ 8.19690}& 0\\
\hline
\end{tabular}
\end{table}

\begin{table}[!t]
\caption{Energies of harmonically-trapped spin-1 anti-ferromagnetic (polar) and ferromagnetic (ferro) quasi-1D and quasi-2D BECs for different values of magnetization $m$. In the quasi-2D case we consider only the circularly-symmetric states. 
 The parameters in 1D: $c_0= 241$, $c_2 = 7.5$ (polar) and $c_0= 885$, $c_2 = -4.1$ (ferro), space step DX = 0.05, time step DT = 0.00025. The parameters in 2D: $c_0= 482$, $c_2 = 15$ (polar) and $c_0= 1327.5$, $c_2 = -6.15$ (ferro), space steps DX = DY= 0.1, time step DT = 0.0005. The SO-coupled (SO-cpld) results in 1D refer to the coupling $\gamma p_x \Sigma_x$ with $\gamma=0.5$ and those in 2D correspond to the Rashba or Dresselhaus coupling with $\gamma =0.5$. For systems with large nonlinearities the numerically obtained energy $E_{\mathrm{num}}$ lies between the TF \cite{TF} and variational (var) limit:
$E_{\mathrm{TF}} < E_{\mathrm{num}} < E_{\mathrm{var}}$. The parameters considered here are appropriate for a Rb and Na BEC with trap parameters of Ref.~\cite{bao}.}
\label{tab4}
\centering
\small
\begin{tabular}{cc|ccccc|cccccc}
\hline
\hline
 & $m$& 1D & 1D & 1D & 1D & 1D& 2D & 2D & 2D & 2D \\ 
 & &spinor &spinor & spinor& spinor & SO-cpld& spinor & spinor & spinor& SO-cpld \\ 
& & & \cite{bao}& var& TF & $\gamma=0.5$& & var & TF&$\gamma=0.5$ \\ 
\hline
& 0& 15.2485 & 15.2485 &15.7522 & 15.2239 &15.1235 &8.3605&8.8155 & 8.2577& 8.1969 \\ 
&0.1 &15.2514 &15.2514 & & &  &8.3617 & & &\\
&0.2 &15.2599 &15.2599 & & &  &8.3652 & & &\\
Polar & 0.3& 15.2743 & 15.2743 & && &8.3710 &&& \\
 &0.4 &15.2945 &15.2945 && & & 8.3793 & &&\\
 & 0.5 & 15.3209 & 15.3209 & && &8.3900 && & \\
&0.6 &15.3537 &15.3537 & & &  &8.4033& &&\\
\hline
 & 0 & 36.1365 & 36.1365 &37.3574& 36.1243&36.0115 &13.7420 & 14.5364 &13.6723 & \\ 
 & 0.0468 & 36.1365 &  & & & &13.7420 & & &13.5976\\
 & 0.1 & 36.1365 & 36.1365 & & & &13.7420 & & &\\
Ferro & 0.2 & 36.1365 & 36.1365 & & & &13.7420 & & &\\
 & 0.3 & 36.1365 & 36.1365 & & & &13.7420& & &\\
 & 0.4 & 36.1365 & 36.1365 & & & &13.7420 & & &\\ 
 & 0.5 & 36.1365 & 36.1365 & & & &13.7420 & & &\\
 & 0.6 & 36.1365 & 36.1365 & & & &13.7420 & & &\\
\hline
\end{tabular}
\end{table}

Before we illustrate our results, we now study in Table~\ref{tab3} the convergence of our calculational scheme in 1D and 2D for SO-coupled ferromagnetic and polar BECs employing the above-mentioned parameters in 1D and 2D upon the reduction of space steps DX and (= DY) from 0.4 to 0.05. In this Table we display the energies, viz.~Eq.~(\ref{energy}), and the 
respective percentage numerical errors for four different sets of parameters in 1D and 2D.
 We see that upon a reduction of space step the energy value 
rapidly converges. The result for energy with space step 0.4 is already very accurate and the result remains unchanged to five significant figures after the decimal point for space steps 0.1 and 0.05 both in 1D and 2D. In the following we will present results of our study with space steps 0.05 and 0.1 in 1D and 2D, respectively. The accuracy increases as the space-step is reduced, but a reduced value of space step requires a larger number of time iterations for convergence. Hence, it is computationally more economic to use a large space step. 

In Table~\ref{tab4} we show the energy, viz. Eq. (1), of an anti-ferromagnetic and a ferromagnetic BEC in quasi-1D and quasi-2D traps for different values of magnetization $m$. For the quasi-2D trap, we consider only the circularly-symmetric states. In 2D the energies are the same for both Rashba and Dresselhaus SO couplings, although the underlying wave functions are different.
 The analytic TF and variational energies are also displayed, for comparison, as well as those from Ref.~\cite{bao}. In the case of a ferromagnetic BEC without SO coupling, the energies are independent of $m$ values, whereas in the anti-ferromagnetic case the dependence on $m$ exists, as can be seen. In both cases the analytic TF and variational energies are independent of $m$. For condensates with large densities as in Table~\ref{tab4}, where the TF results are reliable, the actual energies are larger than the corresponding TF values \cite{TF}. On the other hand, the variational energies are always larger than the actual energies. Hence for large nonlinearities, as we see in Table~\ref{tab4}, the variational and TF energies define the two bounds for the actual, numerically calculated energy. In Table~\ref{tab4} we also present the numerically calculated energies for the SO-coupled BECs in 1D and 2D for $\gamma =0.5$.
 
 \begin{figure}[!b] 
\centering
\includegraphics[width=\linewidth]{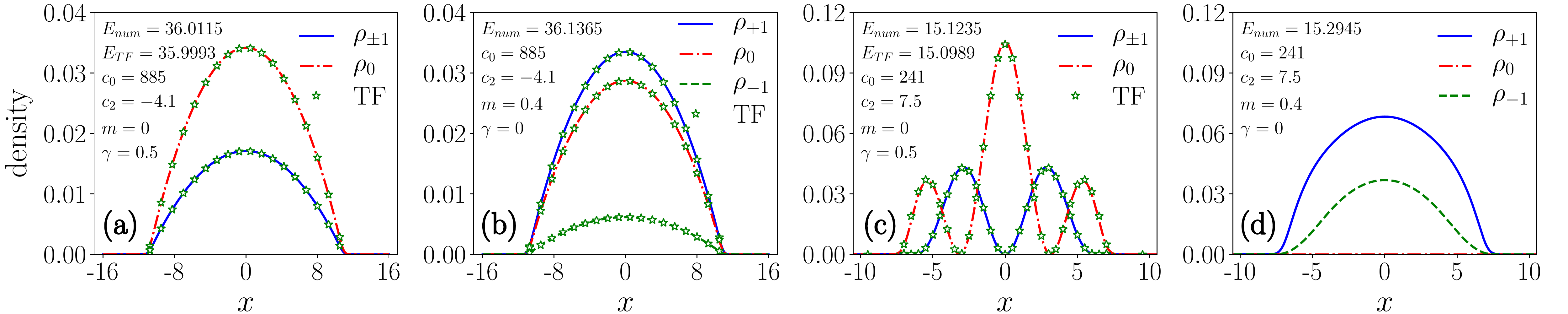} 
\caption{Numerically calculated component density $\rho_j(x)$ (lines) and energy $E$ of a quasi-1D harmonically trapped SO-coupled ferromagnetic BEC with nonlinearities $c_0=885$, $c_2=-4.1$ and  
(a) $\gamma=0.5$ and (b) $m=0.4$, $\gamma=0$, compared with the analytic TF result (chain of symbols). The same for an anti-ferromagnetic BEC with nonlinearities $c_0=241$, $c_2=7.5$ and (c) $\gamma=0.5$ and 
(d) $m=0.4$, $\gamma=0$. The SO-coupling 
is $\gamma p_x \Sigma_x$ with $\gamma =0.5$ in all cases.
All densities are calculated by imaginary-time propagation employing Gaussian input functions. All results reported in this paper are in dimensionless units, as outlined in Section~\ref{sec:GPE}.}
\label{fig2}
\end{figure}

\begin{figure}[!t]
\centering
\includegraphics[width=.6\linewidth]{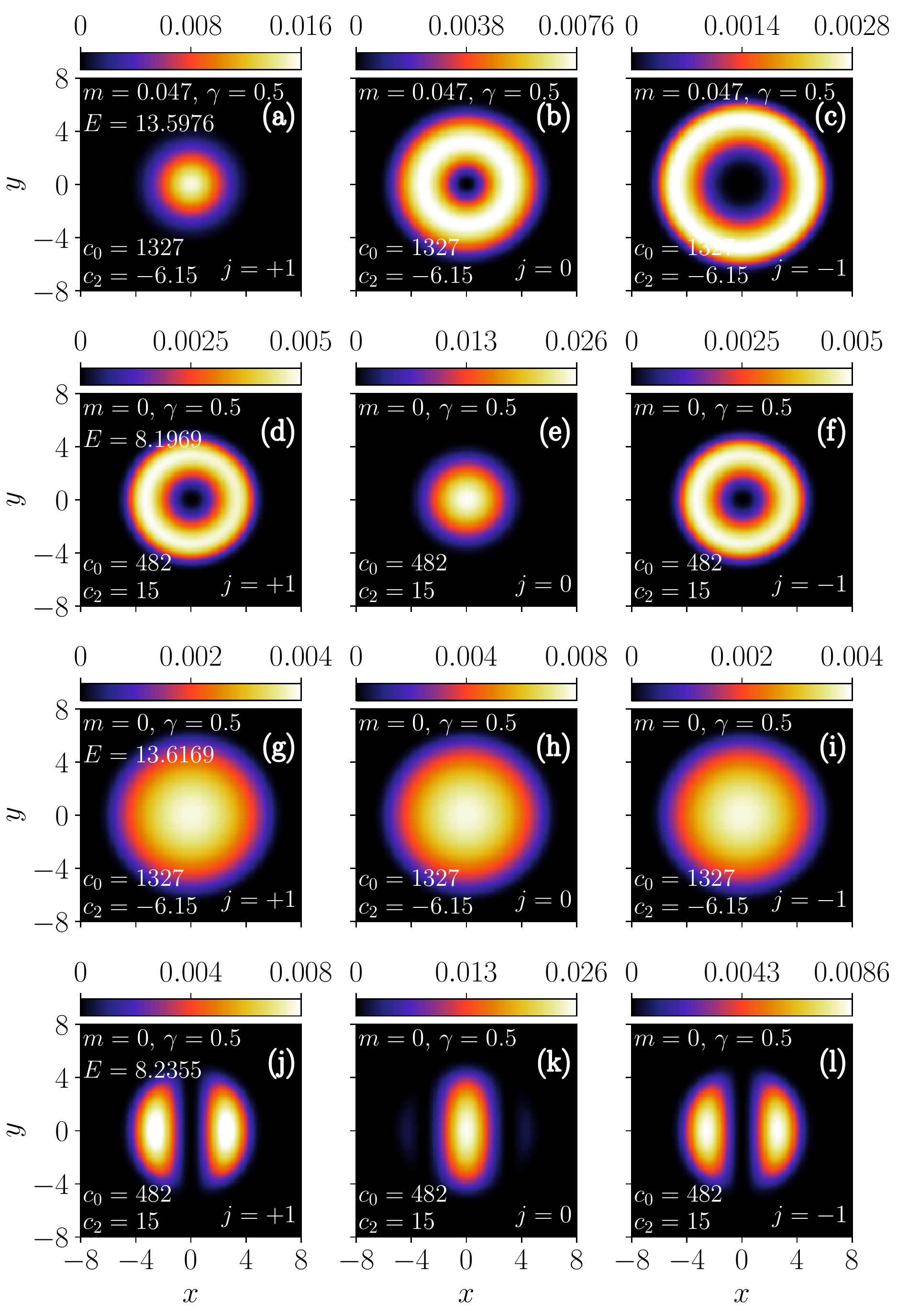}
\caption{Contour plots of numerically calculated component densities $\rho_j(x, y)$ for (a) $j=+1$, (b) $j=0$, and (c) 
$j=-1$ of a quasi-2D harmonically trapped Rashba or Dresselhaus SO-coupled ferromagnetic BEC with nonlinearities $c_0=1327.5$, $c_2=-6.15$, the SO-coupling strength $\gamma=0.5$. 
(d)-(f) A Rashba or Dresselhaus SO-coupled anti-ferromagnetic BEC with nonlinearities $c_0=482$, $c_2=15$, and $\gamma = 0.5$. (g)-(i) A ferromagnetic BEC with $c_0=1327.5$, $c_2=-6.15$, $\gamma=0.5$ for an equal mixture of Rashba and Dresselhaus SO couplings. 
 (j)-(l) An anti-ferromagnetic BEC with $c_0=482$, $c_2=15$, $\gamma=0.5$ for an equal mixture of Rashba and Dresselhaus SO couplings. 
The numerical energies are displayed in plots (a), (d), (g), and (j).} 
\label{fig3}
\end{figure}

 \begin{figure}[!t]
\centering
\includegraphics[width=.6\linewidth]{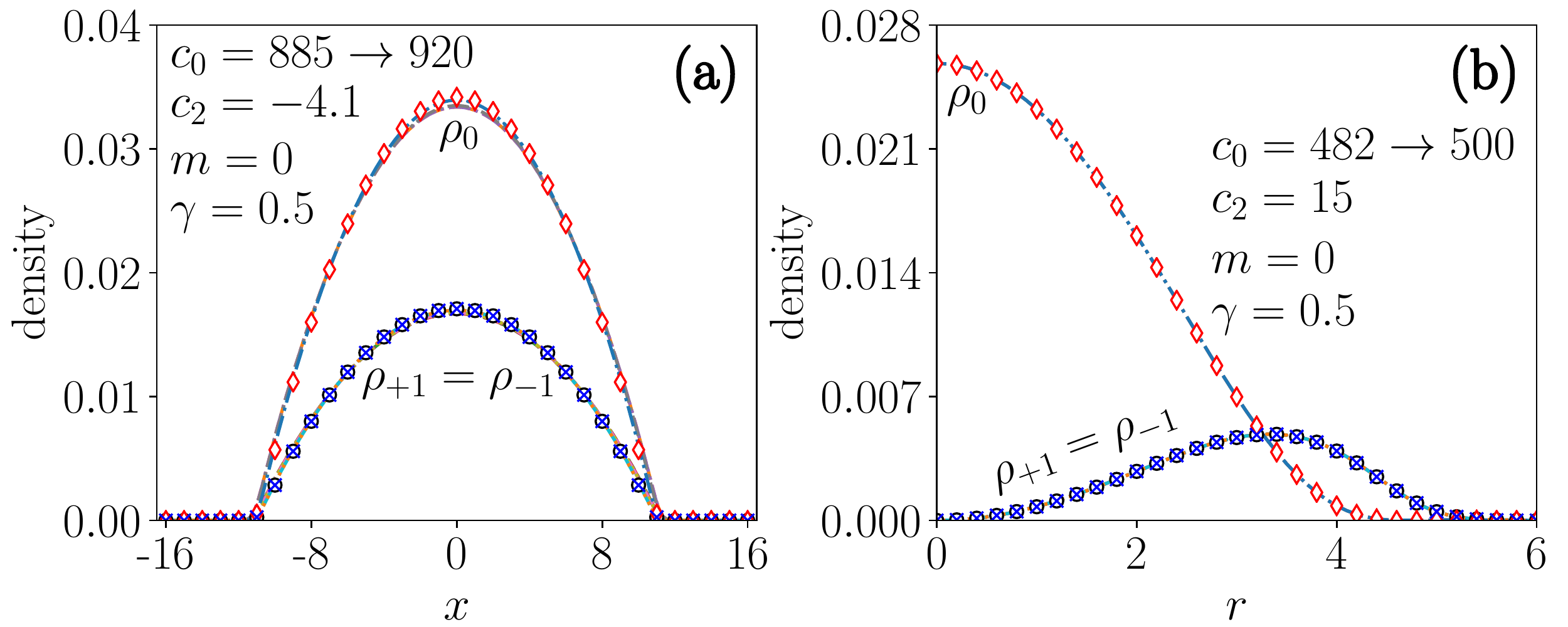}
\caption{(a) Numerically calculated component densities of the quasi-1D SO-coupled ferromagnetic BEC of Fig.~\ref{fig2}(a) with $c_0=885$, $c_2=-4.1$, $\gamma =0.5$ during the real-time evolution at times $t=50,100,150,200$ (full lines), using the imaginary-time wave function as the initial state, compared with the converged imaginary-time densities shown in Fig.~\ref{fig2}(a) (chain of symbols). The parameter $c_0$ was changed from 885 to 920 at $t=0$. (b) Numerically calculated radial component densities of the quasi-2D anti-ferromagnetic Dresselhaus SO-coupled BEC of Fig.~\ref{fig3}(d)-(f) with $c_0=482$, $c_2=15$, $\gamma =0.5$ during the real-time evolution at times $t=50,100,150,200$ (full lines), using the imaginary-time wave function as the initial state, compared with the converged imaginary time densities (chain of symbols). The parameter $c_0$ was changed from 482 to 500 at $t=0$.}
\label{fig4}
\end{figure}

We show the numerically calculated and the TF component densities $\rho_j({\bf x})$ of a quasi-1D harmonically trapped ferromagnetic SO-coupled BEC for $c_0=885$ and $c_2=-4.1$, with $\gamma =0.5$ in Fig.~\ref{fig2}(a) and $m=0.4$, $\gamma =0$ in Fig.~\ref{fig2}(b), along with the corresponding energy values given by Eq. (\ref{energy}). All the states reported here are the lowest-energy ground states for a given set of parameters, obtained by imaginary-time propagation. 
The same quantities are shown for an anti-ferromagnetic BEC with $c_0=241$ and $c_2=7.5$, with $\gamma =0.5$ in Fig.~\ref{fig2}(c) and with $m=0.4$, $\gamma=0$ in Fig.~\ref{fig2}(d). The SO coupling in both cases 
is of type $\gamma p_x\Sigma_x$ with $\gamma =0.5$. These nonlinear parameters $c_0$ and $c_2$ were considered 
previously in Ref.~\cite{bao}. These states are calculated with the analytic initial functions, included by setting NSTP $\ne 0$. 
  
We now exhibit the density of a Rashba SO-coupled quasi-2D spinor BEC for $\gamma=0.5$. 
 The BEC components develop distinct angular momentum structure in this case. To illustrate this, we display in Fig.~\ref{fig3}(a)-(c) the contour plots of densities of a ferromagnetic Rashba or Dresselhaus SO-coupled 
BEC.
In Fig.~\ref{fig3}(d)-(f) the corresponding contour plots of an anti-ferromagnetic Rashba or Dresselhaus SO-coupled 
BEC are shown, while Fig.~\ref{fig3}(g)-(i) display the same for a ferromagnetic 
BEC for an equal mixture of Rashba and Dresselhaus couplings. 
In Fig.~\ref{fig3}(j)-(l) the plots are shown for an anti-ferromagnetic 
BEC, also for an equal mixture of Rashba and Dresselhaus couplings. 
 In all cases the nonlinearities $c_0$ and $c_2$ are the same as in Table~\ref{tab3}. 
 The components $j=+1, 0$ and $-1$ in Fig.~\ref{fig3}(a)-(c) have angular momentum $0,\pm 1,\pm 2$, respectively, for Rashba and Dresselhaus SO couplings. On the other hand, the components in Fig.~\ref{fig3}(d)-(f) have angular momentum $\mp 1,0,\pm 1$, respectively, for Rashba and Dresselhaus SO couplings.
The angular momenta of the spinor components were found from the contour plot of the phases of the corresponding wave functions (not explicitly considered in this paper). 

Finally, we demonstrate the dynamical stability of the imaginary-time results using the real-time propagation for a large interval of time. Using the converged imaginary-time wave function as the initial state, the real-time calculation is initiated after introducing a small perturbation, by changing the value of $c_0$ at $t=0$ by a small amount. First we consider the quasi-1D SO-coupled ferromagnetic spinor BEC of Fig.~\ref{fig2}(a) and perform a real-time simulation for 200 units of time by changing $c_0$ from 885 to 920. The component densities at times $t=50,100,150$, and 200 are plotted in Fig.~\ref{fig4}(a). The imaginary-time converged results (chain of symbols) are also shown, for comparison. Next we consider a quasi-2D anti-ferromagnetic Rashba SO-coupled spinor BEC with $c_0=482, c_2=15, \gamma=0.5$, viz.~Fig.~\ref{fig3}(d)-(f), which is taken as initial state and a real-time propagation is performed for 200 units of time upon changing $c_0$ from 482 to 500 at $t=0$. The radial component densities are plotted at times $t=50,100,150$, and 200 in Fig.~\ref{fig4}(b) together with the converged imaginary-time density (chain of symbols). The fact that all component densities in Fig.~\ref{fig4}(a) and \ref{fig4}(b) for a quasi-1D and a quasi-2D BEC over a large interval of time are stable during the real-time propagation demonstrates the dynamical stability of the condensate.

\section{Summary}
\label{sec:con}

We have presented efficient OpenMP FORTRAN programs for solving the GP equation for a 
three-component spin-1 spinor BEC and used these to calculate the densities and energies for various values of system parameters. Different SO and Rabi coupling terms can be included in the programs. 
We provide two sets of programs: one for a quasi-1D BEC and the other 
for a quasi-2D BEC. Each of these programs is capable of executing both the imaginary- and the real-time propagation. The imaginary-time propagation programs yield appropriate results in agreement 
with variational approximation in all cases \cite{lplga,Alter}. We use the split-step Crank-Nicolson discretization to implement time propagation, relying on our earlier OpenMP FORTRAN programs of Ref.~\cite{bec2017x}. The GP equation can be solved by imaginary- or real-time propagation with an analytic wave function or a pre-calculated numerical wave function as the initial state. 
We stress that the convergence with one initial state could be much faster than with another initial state, and tailoring the input wave function using an analytic or a previously calculated numerical wave function is always an advantage. We have also presented the results for density and energy of different states and compared these with analytic 
variational and TF approximate results. 
 
\section*{Acknowledgments}
\noindent
We thank Dr. Sandeep Gautam for his kind interest and helpful comments.
R.R. acknowledges University Grants Commission (UGC), India for the financial support in the form of UGC-BSR-RFSMS Research Fellowship scheme
(2014-15). D.V. and A.B. acknowledge funding provided by the Institute of Physics Belgrade, through the grant by the Ministry of Education, Science, and Technological Development of the Republic of Serbia. The work of P.M. forms parts of sponsored research projects by Council of Scientific and Industrial Research (CSIR), India under Grant No. 03(1422)/18/EMR-II, and Science and Engineering Research Board (SERB), India under Grant No. CRG/2019/004059.
S.K.A. acknowledges support by the CNPq (Brazil) grant 301324/2019-0, and by the ICTP-SAIFR-FAPESP (Brazil) grant 2016/01343-7.

\appendix 

\section{Detailed Numerical Procedure}
\label{appA}

All terms in Eqs.~(\ref{EQ1}) and (\ref{EQ2}), except the last terms containing explicit complex conjugation, 
have the form of a conventional diagonal GP equation for a three-component BEC, whose solution procedure employing the split-step method is well known. The last non-diagonal terms in Eqs.~(\ref{EQ1}) and (\ref{EQ2}), as well as the Rabi coupling terms proportional to $\widetilde \Omega$ require special attention. In explicit matrix notation, the split-step equation that takes into account the last terms of Eqs.~(\ref{EQ1}) and (\ref{EQ2}), together with the Rabi coupling terms proportional to $\widetilde \Omega$, can be written as 
\begin{align}
i\partial_t \begin{pmatrix}
\psi_{+1} ({\bf r},t)\\
\psi_{0} ({\bf r},t)\\
\psi_{-1}({\bf r},t)
\end{pmatrix}=
\mathbf{A}
\begin{pmatrix}
\psi_{+1} ({\bf r},t)\\
\psi_{0} ({\bf r},t)\\
\psi_{-1}({\bf r},t)
\end{pmatrix}\, ,
\; \quad {\rm where}\quad \mathbf{A} = 
\begin{pmatrix}
0 & a & 0 \\
a^* & 0 &b\\
0 & b^*& 0
\end{pmatrix}\, ,
\label{eq:nu}
\end{align}
and $a=c_2 \psi_{-}^{*} \psi_{0}+\widetilde \Omega$, $b= c_2 \psi_{0}^{*} \psi_{+}+\widetilde \Omega$. The real eigenvalues of the {Hermitian} matrix $\mathbf{A}$ are $\lambda_1=C= \sqrt{|a|^2+|b|^2}$, $\lambda_2=0$, and $\lambda_3= - C$, while the corresponding eigenvectors are $v_1= (v_{11},\, v_{12},\, v_{13})^T = (a,\, C,\, b^*) ^T$, $v_2= (v_{21},\, v_{22},\, v_{23})^T= (b,\, 0,\, -a^* )^T$, and $v_3= (v_{31},\, v_{32},\, v_{33})^T=(a,\, -C,\, b^*) ^T$, respectively. 
For a sufficiently small time step $\Delta$, the solution of Eq.~(\ref{eq:nu}) is given by
\begin{align}\label{eq11}
\begin{pmatrix}
\psi_{+1}({\bf r},t+\Delta) \\
\psi_{0}({\bf r},t+\Delta) \\
\psi_{-1}({\bf r},t+\Delta) 
\end{pmatrix}
=
\mathbf{V}
\begin{pmatrix}
\mathrm{e}^{{- i}\Delta \lambda_1 } & 0 & 0 \\
0 & \mathrm{e}^{{- i}\Delta \lambda_2} & 0 \\
0 & 0 & \mathrm{e}^{{- i}\Delta \lambda_3} 
\end{pmatrix}
{\mathbf{V}^{-1}}
\begin{pmatrix}
\psi_{+1}({\bf r},t) \\
\psi_{0}({\bf r},t)\\
\psi_{-1}({\bf r},t)
\end{pmatrix}\, ,
\end{align}
where
\begin{align}
\mathbf{V}\equiv 
 \begin{pmatrix}
v_{11} & v_{21} & v_{31} \\
v_{12} & v_{22} & v_{32} \\
v_{13} & v_{23} & v_{33} 
\end{pmatrix}=
\begin{pmatrix}
a & b & a \\
C & 0 & -C \\
b^* & -a^* & b^* 
\end{pmatrix}\, ,\quad \quad
\mathbf{V}^{-1} =\frac{1}{2C^2} 
\begin{pmatrix}
a^* & C & b \\
2b^* & 0 & -2a \\
a^* & -C & b 
\end{pmatrix}\, .
\end{align}
In Eq.~(\ref{eq11}) the right-hand-side is considered known, since it is expressed in terms of the wave-function values at time $t$, and thus the wave function is easily propagated to time $t+\Delta$.
In case of larger spins, if the matrix ${\bf A}$ of larger dimension
cannot be analytically diagonalized, 
it can be diagonalized numerically by one of the many available subroutine packages.

When we consider the SO coupling terms proportional to $\widetilde \gamma$, they can be also evaluated at time $t$ using the known wave-function values, and the corresponding split-step equations can be solved to propagate the wave functions to time $t+\Delta$. For example, in case of the $\gamma p_x \Sigma_x$ coupling in 1D, these are performed via 
\begin{align}\label{p1}
\psi_{\pm 1}({\bf r}, t+\Delta) &= \psi_{\pm 1}({\bf r},t) - \Delta \widetilde \gamma \partial_x \psi_0({\bf r},t)\, , \\ \label{p2}
\psi_{0}({\bf r},t+\Delta) &= \psi_{0}({\bf r},t) - \Delta \widetilde \gamma \Big[ \partial_x \psi_{+1}({\bf r},t) + \partial_x \psi_{-1}({\bf r},t) \Big]\, .
\end{align}
In Eqs.~(\ref{p1}) and (\ref{p2}) the right-hand-side at time $t$ is known and hence the wave-function values at time $t+\Delta$ 
can be obtained from those at time $t$. All SO coupling terms are treated in the same fashion.

 The simultaneous maintenance of the normalization and magnetization $m$, viz.~Eq.~(\ref{noma}), during the time propagation, given by Eqs.~(\ref{noma}), is done following the procedure of Ref.~\cite{baonorm}, by rescaling the wave-function components after each time step $\Delta$ according to 
$\psi_j\to d_j \psi_j$, where 
\begin{align}\label{DJ}
d_0=\frac{\sqrt{1-m^2}}{\sqrt{N_0+\sqrt{4(1-m^2)N_{+1}N_{-1}+m^2N_0^2}}}, \quad 
d_1= \frac{\sqrt{1+m-d_0^2N_0} }{\sqrt{2N_{+1}}}, \quad d_{-1}= \frac{\sqrt{1-m-d_0^2N_0} }{\sqrt{2N_{-1}}},
\end{align}
and $N_j=\int d{\bf r}\, \rho_j({\bf r},t)$. However, when the SO coupling does not commute with $\Sigma_z$, we do not impose 
the condition of conservation of magnetization and the rescaling is done according to 
\begin{align}\label{DJ2}
d_0=d_{+1}=d_{-1}=\frac{{1}}{\sqrt{N_0+N_{+1}+N_{-1}}}.
\end{align}

\setcounter{equation}{0}
\renewcommand{\theequation}{S.\arabic{equation}}

\vspace*{1cm}\noindent
{\bf\large Supplementary Material}

\section*{S: Variational and Thomas-Fermi approximations}

The Gross-Pitaevskii Eqs.~(1) and (2) can be derived from the energy functional \cite{thspinor,thspinorb}
\begin{align}\label{energyS}
E=&\frac{1}{2} \int d {\bf r}\, \Bigg\{\sum_j |\nabla _{\bf r}\psi_j|^2+2V({\bf r})\rho + c_0 \rho^2
\nonumber \\ &
+ c_2\left[ \rho_{+1}^2+ \rho_{-1}^2 +2 \left( \rho_{+1}\rho_0+ \rho_{-1}\rho_0- \rho_{+1}\rho_{-1}+\psi_{-1}^*\psi_0^2\psi_{+1}^*+
\psi_{-1}{\psi_0^*}^2\psi_{+1}
\right) 
\right] \nonumber \\ &
+2 \frac{\Omega}{\sqrt 2}\Big[ (\psi^*_{+1}+ \psi^*_{-1})\psi _0+ \psi_0^* (\psi_{+1}+ \psi_{-1}) \Big] 
 + 2
\gamma \Big[ 
\psi_{+1}^*f_{+1}+ \psi_{-1}^*f_{-1} + \psi_0^* g 
 \Big] \Bigg\}.
\end{align}

We describe now two approximation schemes for the solution of the spinor GP 
equations, i.e., the variational approximation and the TF approximation \cite{GA,TF}.
For a ferromagnetic BEC ($c_2<0$) without SO coupling,   the ground-state densities  are essentially proportional to each other, such that
\begin{align}\label{DM}
 \psi_j({\bf r})=\alpha_j \widetilde \psi({\bf r})\,,\, j =\pm 1,0\, ,
\end{align}
where $\alpha_j$ are complex numbers and the function $\widetilde \psi({\bf r})$ is to be determined. If we substitute Eq.~(\ref{DM}) into Eqs.~(1) and (2), we obtain three equations for the unknown function $\widetilde \psi({\bf r})$. For these equations to be consistent, one must have   \cite{GA}
\begin{align} \label{DMEQ}
\mu \widetilde \psi({\bf r})= \left[-\frac{1}{2}\nabla^2+V({\bf r})+g\widetilde \rho({\bf r}) \right] \widetilde \psi({\bf r})\, , \quad
 \widetilde \rho({\bf r}) = |\widetilde \psi({\bf r})|^2\, ,
\end{align}
with $g=(c_0+c_2)$ and subject to  $\int \widetilde \rho({\bf r})\, d {\bf r}=1$, with $\mu$  the chemical potential  and 
$|\alpha_{\pm 1}|=( {1\pm m})/{2}\, , \quad |\alpha_0| = \sqrt{({1-m^2})/{2}}\, .$
For the ground state of an anti-ferromagnetic BEC ($c_2>0$), the $j=0$ component is absent, $\psi_0=0$. In this case Eq.~(\ref{DM}) holds only for $m=0$, with 
$|\alpha_{\pm 1}|=1/\sqrt 2, \alpha_0=0$, while Eq.~(\ref{DMEQ}) is again valid with $g=c_0$.

\section*{S.1: Variational approximation}

Equation (\ref{DMEQ}) can be derived by a minimization of the energy functional  
\begin{align}
 E=\frac{1}{2} \int d {\bf r}\, \Big[\left|\nabla\widetilde \psi({\bf r})\right| ^2
+{\bf r}^2 \left| \widetilde \psi({\bf r})\right|^2+ g\left| \widetilde \psi({\bf r})\right|^4\Big]\, .
\end{align}
We consider the variational ansatz 
 $
\widetilde \psi({\bf r})= \left[{{w\sqrt{\pi}}} \right]^{-d/2}\exp\left(-\frac{{\bf r}^2}{2w^2}\right)\,
 $
where $d=1$ in 1D and $d=2$ in 2D, and $w$ is a variational parameter. With this ansatz the energy functional becomes
\begin{align}\label{energyvar}
E=\frac{d}{4 w^2}+ \frac{d w^2}{4} + \frac{g}{2\left(w\sqrt{2\pi}\right)^d}\, .
\end{align}
The parameter $w$ is obtained by a minimization of this energy.  

 \section*{S.2: Thomas-Fermi approximation}

In absence of SO coupling ($\gamma =0$), for an anti-ferromagnetic ($c_2>0$)  spin-1 BEC  with zero magnetization and for a  ferromagnetic ($c_2<0$) spin-1 BEC, there are simple analytic solutions 
based on a decoupled mode TF approximation and we quote the results 
here. The density of the three components are given by \cite{GA}
\begin{align}
\rho_j({\bf r})= |\alpha_j|^2 \rho_{\mathrm{TF}}({\bf r}) = |\alpha_j|^2 \frac{ L_d^2-{\bf r}^2} {2g}\, 
\end{align} 
where $L_1=[3g/2]^{1/3}$ in 1D and $L_2=[4g/\pi]^{1/4}$ in 2D. In 1D this density is the linear density, while in 2D it is the radial density. The TF energy
can be evaluated and yield $E_{\mathrm{TF}}=3[3g/2]^{2/3}/10$ in 1D, and $E_{\mathrm{TF}}=2\sqrt{g/\pi}/3$ in 2D. 

Some useful analytic results can be obtained for a ferromagnetic BEC in 1D in the absence of magnetization ($m=0$) for non-zero SO ($\gamma \ne 0$) and Rabi ($\Omega \ne 0$) coupling strengths. The ansatz 
for the wave function components is taken to be $\alpha_{\pm 1}=e^{i \gamma x}/2, \alpha_0 = - e^{i \gamma x}/\sqrt{2}$ in Eq.~(\ref{DM}), viz.~Eq.~(22) of Ref.~\cite{lplga}. For the SO coupling of the form $\gamma p_x \Sigma_x$, the total energy for small values of $\gamma \ne 0$   
can be now evaluated to yield
$\widetilde E _{\mathrm{TF}}= E_{\mathrm{TF}}- \frac{\gamma^2}{2} - \Omega\, ,$
where $E_{\mathrm{TF}}$ are the above calculated expressions in 1D or 2D. In this case the TF profile for the density remains unchanged for small $\gamma$ and $\Omega$. However, the energy gets modified due to the SO and Rabi couplings.  
 
Additional useful analytic results can be obtained for an anti-ferromagnetic BEC in 1D in the absence of magnetization ($m=0$) for a non-zero SO coupling strength ($\gamma \ne 0$). The ansatz 
for the wave function components in this case is taken to be
$\alpha_{\pm 1}=i\sin(\gamma x)/\sqrt{2}$, $\alpha_0 = - \cos(\gamma x)$ in Eq.~(\ref{DM}), viz.~Eq.~(18) of Ref.~\cite{lplga}. For the SO coupling of the form $\gamma p_x \Sigma_x$, the total TF
energy for small values of $\gamma \ne 0$ and $\Omega \ne 0$ can be now evaluated: 
$\widetilde E_{\mathrm{TF}}= E_{\mathrm{TF}}- \frac{\gamma^2}{2}\, .$
The TF profile for the density remains unchanged for small $\gamma$, but the energy gets modified due to the SO coupling.

\end{document}